\documentclass[fleqn,twoside]{article}
\usepackage{amsmath,espcrc2,graphicx,epsfig}

\title{Borel resummation with low order perturbations in QCD}

\author{Taekoon Lee\address{Department of Physics, Kunsan National University,
Kunsan 573-701, Korea}%
        \thanks{This work was supported in part by the 
Korea Research Foundation 
Grant (KRF-2005-015-C00107) and
research fund from Kunsan National University.}}

\begin{document}

\begin{abstract}
The bilocal expansion of Borel transform
provides an efficient way of Borel resummation with low order perturbations in QCD.
Its applications to the heavy quark  pole mass, static potential, and
lattice calculation are reviewed.
\end{abstract}

\maketitle

\section{Introduction}
The perturbative expansions in weak coupling constant in quantum field theories
are in general divergent, with the coefficients growing in factorial.
One way to make a rigorous sense out of the divergent series is to resum the
series using Borel transform. When the large order behavior of the expansion
is sign-alternating Borel resummation may yield the full
amplitude. But in QCD the series are in general of same-sign and not Borel
resummable, and must be augmented by nonperturbative effects.
However, Borel resummation can still be useful in these cases when 
a precise definition of the nonperturbative effects are required, or 
the coupling is large enough that the usual perturbative amplitude 
from the low order perturbations is not sufficiently accurate for the purpose.

The same-sign large order behavior of weak coupling expansion gives a
singularity (renormalon) on the contour of the Borel integral. When the coupling is small
the Borel integral receives most of its contribution from the immediate neighborhood
of the origin, where the Borel transform can be well-described by the low order perturbations,
and the singular behavior of the Borel transform is not
of much concern, but as the coupling increases the contribution from
the region near the singularity becomes important, and one needs to have an
accurate description of the Borel transform in the region that contains the
origin as well as the nearest singularity to the origin.

It is obvious that it is a formidable task to rebuild the singular behavior of
the Borel transform using the ordinary perturbative expansions about the origin.
In fact, the solvable models like the quantum mechanical instanton models
suggest that one need many 10s orders of perturbative terms  to obtain a
good Borel resummation. This approach is obviously not viable in QCD
considering the high cost of loop calculations. Thus an alternative approach is
required to do any attempt for realistic Borel resummation with the low order
perturbations in QCD.

The nature of the renormalon singularity can be determined by
renormalization group technique, and the (singular part of) 
Borel transform near the singularity can 
be systematically expanded perturbatively, up to the residue of  the
singularity that determines the overall normalization constant of the large
order behavior (For a review see \cite{beneke}). 
It was later realized that, once the nature of the
singularity is known,  the residue too can
be calculated perturbatively,
using  only the usual perturbative expansions 
of the quantity in concern
\cite{lee.resi}.

We thus have a systematic expansion of the Borel transform at two points,
the origin as well as the singularity. The bilocal expansion is to combine
these expansions via an interpolation to obtain an accurate description of the
Borel transform not only about the origin but also about 
the singularity
\cite{lee.bilocal}.

The bilocal expansion has been successfully applied to several QCD
observables \cite{lee.bilocal,apps,apps.gardi},
and following is a review of those applications.

\section{Bilocal expansion}
Let us assume that the Borel transform $\tilde A(b)$ of an amplitude has
the perturbative expansion about the origin to order $\rm M$ as
\begin{eqnarray}
\tilde A(b)= \sum_{i=0}^{M} \frac{a_i}{i!} b^i 
\label{exp.org}
\end{eqnarray}
and expansion about the nearest renormalon singularity to order $\rm N$ as
\begin{eqnarray}
\tilde A(b)&= &\frac{\cal N}{(1-b/b_0)^{1+\nu}}\left[ 1+\sum_{j=1}^{N} c_j
(1-b/b_0)^j\right] \nonumber \\
&&+{\rm Analytic \,\,\,\,part}
\label{exp.sing}
\end{eqnarray}
where $b_0$ is the location of the singularity. The ``Analytic part'' denotes
terms that are analytic on the disk $|b-b_0|<b_0$, and in general is not known.

We can now combine the expansions (\ref{exp.org}) and (\ref{exp.sing})
as a bilocal expansion using the following
interpolation \cite{lee.bilocal}:
\begin{eqnarray}
\tilde A_{\rm M,N}(b)&=& \sum_{i=0}^{M} \frac{h_i}{i!} b^i + \nonumber \\
&&\hspace{-1.5cm}\frac{\cal N}{(1-b/b_0)^{1+\nu}}\left[ 1+\sum_{j=1}^{N} c_j
(1-b/b_0)^j\right] 
\label{bilocal}
\end{eqnarray}
When we compare this expression to (\ref{exp.sing}) we see that the first
sum simulates the ``Analytic part''. The coefficients $h_i$ can be determined
by demanding that the bilocal expansion, when expanded about the origin,
reproduce the perturbative expansion (\ref{exp.org}).
This yields the relation
\begin{eqnarray}
h_0&=& a_0 -{\cal N}(1+\sum_{j=0}^N c_j) \,, \nonumber\\
h_1&=&a_1 -\frac{{\cal N}}{b_0}[1+\nu +\sum_{j=1}^Nc_j (1+\nu-j)] \,,\nonumber\\
h_2&=& a_2-\frac{{\cal N}}{b_0^2}[(\nu+1)(\nu+2)+\sum_{j=1}^N[c_j
(j(j-1) \nonumber\\
&& -2(1+\nu)j+(\nu+1)(\nu+2))]\,,  
\mbox{\small etc.}
\label{h-coeffs}
\end{eqnarray}
The exact Borel transform is given by the limit of the bilocal expansion:
\begin{eqnarray}
\tilde A(b)= \lim_{{\rm M,N}\to\infty} \tilde A_{\rm M,N}(b)
\end{eqnarray}

For the bilocal expansion to be usable it is necessary to compute the residue
$\cal N$. Since the residue determines the normalization
constant of the large order behavior, the Borel transform in bilocal
expansion (\ref{bilocal}) allows one to resum  the divergent series to
all orders once the residue is known.

The residue can be perturbatively computed as follows. Noticing that
\begin{eqnarray}
{\cal N}= R(b_0)
\end{eqnarray}
where
\begin{eqnarray}
R(b)=\tilde A(b) (1-b/b_0)^{1+\nu}
\end{eqnarray}
one can expand R(b) about the origin to evaluate ${\cal N}$. Notice that the
expansion involves only the usual perturbative coefficients $a_i$. 
Although $R(b)$ is not analytic at $b_0$, the expansion
is nevertheless convergent because $R(b)$ is bounded at the singularity.
Fortunately, for several QCD observables this expansion gives a 
rather quick convergence. Adler function is such an example (see Table
\ref{adler}) \cite{lee.resi}.
\begin{table}[t]
\caption{\protect\small\it Residue for the 
Adler function with $N_f$ quark flavor.}
\vspace*{0.3cm}
\label{adler}
\begin{tabular}{ccccc}
\hline
  & $N_f=1$ & $N_f=2$ & $N_f=3$ & $N_f=4$  \\
  \hline
LO &0.904 &0.904 &0.946 & 1.018  \\
NLO &0.521 &0.546 &0.592 & 0.674 \\
NNLO &0.592 &0.549 &0.494 & 0.411 \\
\hline
\end{tabular}
\end{table}
\section{Pole mass}
The perturbative coefficients for the pole mass grow rapidly, and the
convergence for the bottom quark, for example, is not good, 
as can be seen in
\begin{eqnarray}
m_{\rm pole}&=&m_{\overline{\mbox{\small MS}}}( 1+0.093+0.045+0.032)\,\,\,
\end{eqnarray}
at $\alpha_s=0.22\,,\,\,\,N_f=4$.

The Borel transform for the heavy-quark pole mass has the nearest renormalon
singularity at $b=1/2$ and the bilocal expansion can be computed up to the
order ${\rm M=2, N=2}$. The perturbative calculation of the residue shows an
excellent convergence \cite{lee.bilocal,pineda}
\begin{eqnarray}
{\cal N}=0.4244+0.1224+0.0101=0.5569 \,.
\label{residue}
\end{eqnarray}
With this residue and  $\rm N=2$,  and varying $\rm M$ from 0 to 2 we obtain
the Borel resummed bottom quark pole mass $m_{\rm BR}$ as
\begin{eqnarray}
m_{\rm BR}/m_{\overline{\mbox{\small MS}}}=
1+ 0.1577 +0.0041-0.0003 \,.
\end{eqnarray}
Notice the remarkable convergence. The leading term already contains the bulk of the
contribution, which shows that the proper handling of the singularity in the
bilocal expansion accelerates the convergence in an enormous way. 

As the above calculation shows the Borel resummation allows one to have a 
precise relation between the pole mass and $\overline{\rm MS}$ mass.
This means that the pole mass defined through Borel resummation is as
physical as the  $\overline{\rm MS}$ mass and can be used as an alternative
to the short distance masses. 

\section{Static potential}
\begin{figure}[htb]
\epsfig{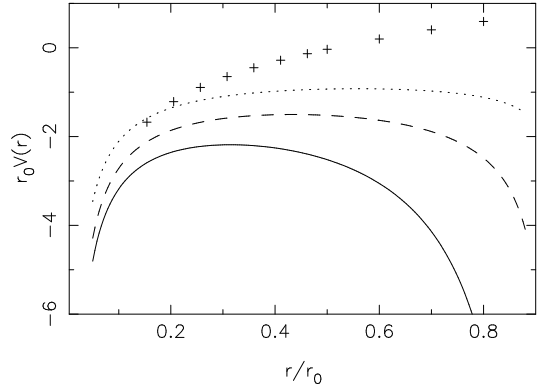}
\vspace*{-0.9cm}
\caption{\protect\small\it 
The static potential at leading order (dotted), next-leading order (dashed),
and next-next-leading order (solid).
The data points denote lattice potential.}
\label{s.pot.PT}
\end{figure}

The  heavy quark static potential obtained in perturbation theory disagrees
badly with the lattice simulation as shown in Fig.\ref{s.pot.PT}, even
in short distances where perturbation is expected to work.

\begin{figure}[htb]
\epsfig{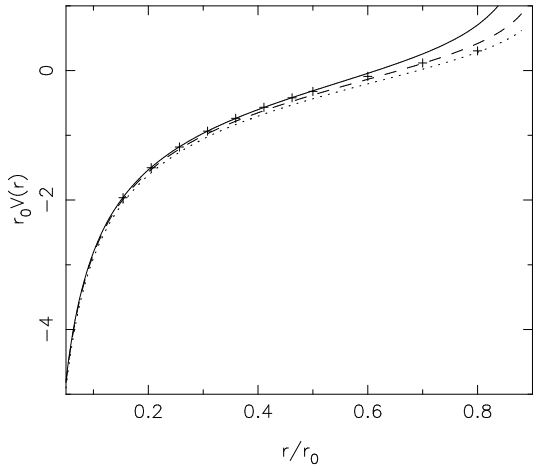}
\vspace*{-0.9cm}
\caption{\protect\small\it Borel resummed potential}
\label{s.pot.BR}
\end{figure}

The nature of the nearest renormalon singularity is the same as the pole mass
and the Borel resummation can be performed in a similar fashion
\cite{lee.bilocal}.
The result
is given in Fig.\ref{s.pot.BR}. The resummed potential agrees
remarkably well with the lattice computation. This is a concrete example
that shows
the efficiency of the bilocal expansion.

\section{Lattice determination of heavy-quark mass}
 The matching relation for
the heavy-quark mass (b-quark) in  lattice 
heavy quark effective theory  reads~~\cite{mar-sac}
\begin{equation}
m_{\rm b}^{\rm pole} =M_{\rm B} - \bar\Lambda \,, 
\quad \bar\Lambda\equiv
{\cal E}(a) -\delta m(a) \,,
\label{rel1}
\end{equation}
where $\bar\Lambda$ denotes the renormalized binding energy, 
$a$ is the lattice spacing, $M_{\rm B}$ is the B-meson mass,
$\delta m$ is the
mass shift in the static limit, and ${\cal E}$ is the binding energy
that is to be computed in lattice simulation. 

The matching relation suggests that with high order perturbative calculations
of the pole mass and the mass shift an accurate determination of the
$\overline{\rm MS}$
mass be possible with precision calculation of $\cal E$. There is, however, a
problem with that the pole mass and mass shift do not converge well perturbatively,
due to the renormalon. To overcome this difficulty the matching relation is
expressed in terms of the $\overline{\rm MS}$ mass only, bypassing the pole mass,
as 
\begin{eqnarray}
m_{\rm b}^{\overline{\rm MS}}&=&\Delta(a)
\large(1+ \nonumber \\
&&\,\,\,\,\sum_{n=0}^\infty r_n(\mu/m_{\rm b}^{\overline{\rm MS}},\mu a)
\alpha_s(\mu)^{n+1}\large)\,,
\label{pertmethod}
\end{eqnarray}
where
$\Delta(a)\equiv M_{\rm B} -{\cal E}(a)$.

Although this expression does not suffer from the renormalon divergence
there could be a convergence problem coming from the mixing of the two
independent scales, the lattice spacing and the heavy quark mass, through the
RG scale $\mu$. When the two scales are close, as in the bottom quark case,
this problem may not be severe but when the two scales are far-separated
this may become a serious problem.

The scale mixing problem can be solved with Borel resummation.
The matching relation (\ref{rel1}) with Borel resummation reads
\begin{eqnarray}
m_{\rm b}^{\rm BR} &=&M_{\rm B} - \bar\Lambda^{\rm BR}\,,
\label{BRmatching}
\end{eqnarray}
where 
$\bar\Lambda^{\rm BR}\equiv {\cal E}(a) -\delta m^{\rm BR}(a)$.
Here the pole mass and mass shift are resummed independently, using
the strong coupling at a scale optimal for each quantities. The scales
are cleanly isolated, with $1/a$ only in $\delta m^{\rm BR}$ and 
$m_{\rm b} $ only in  $m_{\rm b}^{\rm BR}$.

Table ~\ref{table2} shows the $\overline{\rm MS}$ mass for the bottom quark
from the above two approaches in quenched limit. In this case the results from the
two approaches agree remarkably well.
\begin{table}[b]
\caption{\protect\small\it The $\overline {\rm MS}$ mass 
$m_{\rm b}^{\overline {\rm MS}}$ determined in the Borel
summation method and PCM. The units are in GeV.}
\vspace*{0.3cm}
\label{table2}
\begin{tabular}{l|ccc} \hline
$a^{-1}$            &2.12    &2.91    & 3.85    \\ \hline
BR method           & 4.312 &4.312  &4.297   \\ 
PT method &4.319  &4.320  &4.311   \\ \hline
\end{tabular}
\end{table}
 However, when the lattice spacing and the heavy-quark mass are far-different
the scale mixing problem appears. For an (imaginary) 
heavy-light meson with mass $m_H=500$ GeV we obtain, using (\ref{BRmatching}),
the BR mass:
\begin{equation}
m_{\rm Q}^{\rm BR}= 499.542\,, 499.542\,, 499.523\, {\rm (GeV)}\,
\label{Q-brmass}
\end{equation}
at $1/a=$ 2.12, 2.91, 3.85 (GeV), respectively,
and using the relation
\begin{equation}
m_{\rm Q}^{\rm BR}/m_{\rm Q}^{\overline{\rm
MS}}=1+0.03638+0.00008-0.00004
\label{ratio1}
\end{equation}
we obtain the corresponding $\overline {\rm MS}$ 
mass
\begin{equation}
 m_{\rm Q}^{\overline {\rm MS}} =
 481.984\,,481.984\,,481.965 \,{\rm (GeV)}\,.
\label{msmass1}
\end{equation} 
What is remarkable with this result is that the uncertainties in these
numbers are only $\pm 20$ MeV,  entirely coming from  $\bar\Lambda^{\rm BR}$
and the conversion to $ m_{\rm Q}^{\overline {\rm MS}}$ from the BR mass, both
of which cause less than 20 MeV uncertainty. 

On the other hand, with the conventional method using (\ref{pertmethod}),
the $ m_{\rm Q}^{\overline {\rm MS}}$ obtained shows strong dependence
on the RG scale $\mu$ which can be anywhere in a wide range between $1/a$ and
$M_{\rm Q}$, and  the uncertaintities
in the heavy quark mass from this approach
are several hundred MeVs, which are an order of magnitude larger than those
in the Borel resummation method. For details we refer the readers 
to \cite{lee.lattice}.

\section{Conclusion}
With the bilocal expansion  Borel resummation for certain QCD observables
can be done  to a remarkable accuracy with
the low order perturbations. The salient feature of the Borel resummation is
the automatic
isolation of hierarchical scales, which can be useful in quarkonium
systems.

\end{document}